\documentclass[prl,twocolumn,super,showpacs,groupedaddress,superscriptaddress]{revtex4}
\usepackage{graphicx}
\usepackage{color}
\usepackage{amsmath}
\usepackage{latexsym}
\usepackage{amssymb}

\begin{document}
\title{Active polymer translocation through flickering pores}

\author{Jack A. Cohen}
\email[]{j.cohen@physics.ox.ac.uk}
\affiliation{Rudolf Peierls Centre for Theoretical Physics, University of Oxford, Oxford OX1 3NP, UK}

\author{Abhishek Chaudhuri}
\email[]{a.chaudhuri1@physics.ox.ac.uk}
\affiliation{Rudolf Peierls Centre for Theoretical Physics, University of Oxford, Oxford OX1 3NP, UK}
\affiliation{Department of Biomedical Science, University of Sheffield, Sheffield S10 2TN, UK}

\author{Ramin Golestanian}
\email[]{ramin.golestanian@physics.ox.ac.uk}
\affiliation{Rudolf Peierls Centre for Theoretical Physics, University of Oxford, Oxford OX1 3NP, UK}

\date{\today}

\begin{abstract}
Single file translocation of a homopolymer through an active channel under the presence
of a driving force is studied using Langevin dynamics simulation. It is shown that
a channel with sticky walls and oscillating width could lead to significantly more
efficient translocation as compared to a static channel that has a width
equal to the mean width of the oscillating pore. The gain in translocation exhibits
a strong dependence on the stickiness of the pore, which could allow the polymer
translocation process to be highly selective.
\end{abstract}

\maketitle

{\em Introduction.}---The translocation of a polymer through a pore
is important in the context of many biological processes such as
the transport of RNA through a nuclear membrane pore
\cite{salman} and the injection of viruses.
Its various technological applications such as drug delivery \cite{meller1},
rapid DNA sequencing \cite{meller1,meller2,kasianowicz}
and gene therapy has led to several recent experimental
\cite{kasianowicz,akeson,sauer,storm} and theoretical studies \cite{sung,
muthukumar,lubensky,metzler,kantor,milchev,luo1,luo2,sakaue1,gerland,slonkina,
gopinathan,fazli}. Most theoretical work in this field
has focussed on the underlying physics of the translocation process
\cite{sung,muthukumar,kantor} and
the effects of the pore-polymer interactions \cite{luo1,lubensky,slonkina},
structure of the pore \cite{gerland}, crowding \cite{gopinathan}
and confinement effects on the dynamics of translocation. Experimental studies
on electric field driven translocation of DNA and RNA molecules across
$\alpha$-hemolysin channels \cite{kasianowicz} prompted the
introduction of additional driving forces to aid translocation. In most of these
studies the pore is considered static with the polymer always experiencing
a constant confinement during its translocation from the {\it cis} to the
{\it trans} side. However, there are a number of biological examples,
such as the twin-pore translocase complex in the inner membrane of mitochondria
\cite{mitochondria} and the nuclear pore complex \cite{npc}, where
it is known that the width of the channel effectively changes during
the course of translocation. Inspired by these examples, we set out to study
the generic effect of such temporal modulations on the efficiency
of polymer translocation, using a simple coarse-grained model.

We find that the translocation of a polymer through a narrow channel
with a width that oscillates with a given frequency can be significantly
enhanced when compared to a static pore. The time of translocation
is sensitive to the initial condition, and the driving force and
stickiness of the channel walls significantly affect the translocation
at the limit of high frequencies.

\begin{figure*}
\includegraphics[width=1.0\linewidth]{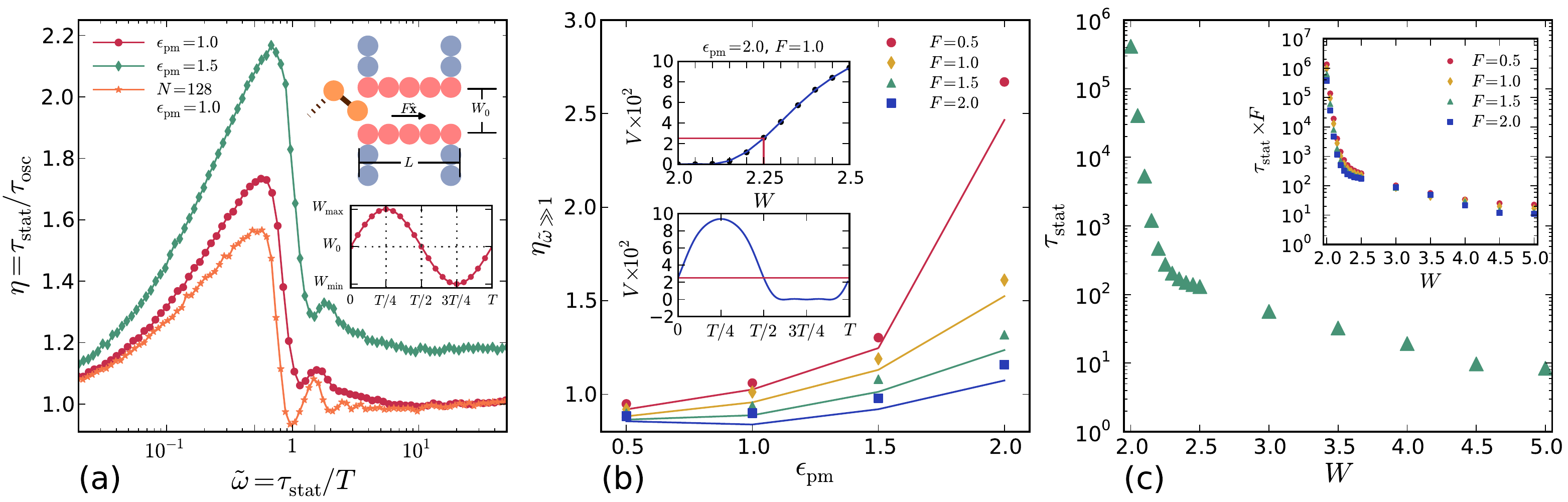}
\caption{\label{f:sr} (a) Gain as a function of the scaled frequency, for $F = 1.0$.
Top inset is a schematic of the simulation geometry. Bottom inset shows an oscillation
cycle of the pore. (b) High frequency limit of the gain for different values
of $\epsilon_\mathrm{pm}$ and $F$. The points are simulation data for oscillating
widths and the lines are extracted from an instantaneous static width approximation
(see text). Top inset shows the translocation velocity as a function of the static width.
Bottom inset shows the velocity as a function of time during the oscillation cycle.
(c) The translocation time as a function of width for static pores,
for $\epsilon_\mathrm{pm} = 1.5$ and $F = 1.5$. The points are
simulation data. Inset shows the same plot for different forces.}
\end{figure*}

{\em Model.}---In our simulation, we model the polymer as a bead-spring chain.
The polymer beads experience an excluded volume interaction modeled by a
repulsive Lennard-Jones (LJ) potential of the form
$U_\mathrm{mm}^\mathrm{LJ}(r) =
4\epsilon \left[ \left(\frac{\sigma}{r} \right)^{12} -
\left(\frac{\sigma}{r}\right)^{6} \right]  + \epsilon$ with a cut-off
at $r = 2^{1/6}\sigma$, where $\sigma$ is the diameter of a bead and
$\epsilon$ gives the strength of the LJ potential. The monomers of the chain
experience an additional interaction modeled as a finitely extensible nonlinear elastic 
(FENE) spring: $U_\mathrm{ch}^\mathrm{FENE}(r) =
-\frac{1}{2} k R^2 \mathrm{ln} \left(1-\frac{r^2}{R^2} \right)$ where
$k$ is the spring constant and $R$ is the maximum separation between
consecutive monomers along the chain. We consider a two-dimensional
(2D) geometry where the pore is modeled as being made up of
monomers of diameter $\sigma$ as shown in Fig. \ref{f:sr}a (inset).
The walls perpendicular to the pore are
taken to be long enough to avoid the crossing of the polymer ends.
The interaction of the polymer with the wall, $U_\mathrm{wm}^\mathrm{LJ}$,
is modeled by the same excluded volume interaction that exists between the polymer beads;
thus, $U_\mathrm{wm}^\mathrm{LJ} = U_\mathrm{mm}^\mathrm{LJ}$. The interaction of
the polymer with the pore is modeled by the Lennard-Jones potential
$U_\mathrm{pm}^\mathrm{LJ}(r) = 4\epsilon_\mathrm{pm} \left[
\left(\frac{\sigma}{r} \right)^{12} - \left(\frac{\sigma}{r}\right)^{6}
\right]$ for $r \le 2.5\sigma$ and $0$ for $r > 2.5\sigma$.
Inside the pore, the polymer experiences an external driving force
$\mathbf{F}_\mathrm{ext} = F \hat{\mathbf{x}}$ directed along the pore axis.
A Langevin dynamics algorithm is used to integrate the
equation of motion of the polymer beads
\begin{math}
m{\bf \ddot{r}}_i=- {\mbox{\boldmath$\nabla$}} U_i + \mathbf{F}_\mathrm{ext} - \zeta {\bf v}_i + {\mbox{\boldmath$\eta$}}_{i},
\end{math}
where $m$ is the monomer mass,
$U_i = U_\mathrm{mm}^\mathrm{LJ} + U_\mathrm{ch}^\mathrm{FENE} + U_\mathrm{wm}^\mathrm{LJ} + U_\mathrm{pm}^\mathrm{LJ}$
is the total potential experienced by a bead, $\zeta$ is the friction coefficient, ${\bf v}_i$
is the monomer velocity, and ${\mbox{\boldmath$\eta$}}_i$ is the random force
satisfying the fluctuation--dissipation theorem
$\langle {\mbox{\boldmath$\eta$}}_{i}(t) \cdot {\mbox{\boldmath$\eta$}}_{j}(t_0) \rangle
=4 k_{\rm B} T \zeta \delta_{ij}\delta (t-t_0)$.

In our model, $\epsilon$, $\sigma$, and $m$ set the units of energy, length,
and mass, respectively, which result in a unit of time as $(m\sigma^2/\epsilon)^{1/2}$.
Using these units, the dimensionless parameters of $R = 2$, $k = 7$, $k_{\rm B}T = 1.2$,
and $\zeta = 0.7$ have been chosen for the simulations, in accordance with earlier simulation
studies of polymer translocation dynamics \cite{luo1,luo2}. The length of
the pore is fixed at $L = 5$, and the length of the polymer is $N = 32$ unless otherwise
specified. We consider the case where the width of the pore,
$W(t)$, is allowed to oscillate harmonically, with frequency $\omega=2 \pi/T$
and amplitude $W_A$, about an average width $W_0={\langle W(t) \rangle}_T$, namely
\begin{math}
W(t) = W_0 + W_A \sin \left(\omega t + \phi\right),
\end{math}
where $\phi$ is an initial phase. Note that $W_\mathrm{min}=W_0 - W_A$ and
$W_\mathrm{max} = W_0 + W_A$ are the minimal and maximal widths of the pore, respectively.
Due to the strongly repulsive excluded volume interaction between beads,
reducing the pore width below a certain value could result in the breaking of
the bonds between neighboring monomers in the polymer. We choose
$W_\mathrm{min}$ such that in a static pore with this width a polymer would be
trapped without breaking up. Moreover, the requirement of single file translocation
of the polymer (i.e. no hairpin structures), would limit $W_\mathrm{max}$.
In accordance with these restrictions, we choose $W_0 = 2.25$ and
$W_A = 0.25$, so that the pore oscillates between $W_\mathrm{min} = 2.0$ and
$W_\mathrm{max} = 2.5$. (The value for the relative change in width might appear too large if regarded
as a conformational change in a protein. However, this is inflicted by the particular model
of rigid spheres used in our coarse-graining, and could be much smaller for more realistic
descriptions.) We have checked that our results are not sensitive to the particular value
of the amplitude, so long as it satisfies the requirements that it closes the channel
at minimum width and allows relatively free passage of the polymer at maximum width.
The driving force inside the pore varies between
$0.5$ and $2$. Initially, the first monomer of the polymer is held fixed at
the entrance of the pore while the other beads are allowed to fluctuate.
After allowing sufficient time for the polymer configuration to equilibrate,
the first monomer is released and the time that elapses between the entrance
of this monomer into the pore and the exit of the last monomer is measured.
This gives the translocation time of the polymer through the pore.
The time step in our simulations is chosen as $\Delta t = 0.01$ and the
averaging is done over 2000 successful translocation events.

{\em Results.}---We examine the efficiency of the translocation process
by comparing the average time of translocation for the oscillating pore
$\tau_\mathrm{osc}$ with the average translocation time for a static pore
(of width $W_0$) $\tau_\mathrm{stat}$. We first focus on the $\phi=0$ case.
In Fig. \ref{f:sr}a, the gain in translocation rate defined as
$\eta \equiv \tau_\mathrm{stat}/\tau_\mathrm{osc}$ is plotted as a function of
the dimensionless frequency $\tilde{\omega}\equiv\tau_\mathrm{stat}/T$.
We find that the translocation rate is enhanced for the oscillating
pore as compared to the static pore. For the $N=32$ case, there are two distinct peaks
at $\tilde{\omega} \sim 0.5$ and $1.5$, with the translocation time at
$\tilde{\omega} \sim 0.5$ almost half of that for the static pore.
This behavior can be understood by noting that during the first half period
of oscillation of the pore (between $t = 0$ and $t = T/2$), the width of
the pore is always greater than the average width, $W_0$, which is
the width of the static pore. When the oscillation frequency is sufficiently
small, then the polymer only experiences this half of the cycle before
the completion of the translocation process. The average translocation time
therefore increases steadily for small frequencies. Beyond the critical
oscillation frequency corresponding to $\tilde{\omega} \sim 0.5$, the
polymer experiences the effect of the second half of the cycle where the
pore width is always smaller than the static pore width, $W_0$.
The average translocation time increases, thus lowering the gain
substantially, until it reaches a minimum corresponding to $\tilde{\omega} \sim 1$.
For higher oscillation frequencies, the polymer starts experiencing the next
half of the cycle between $t = T$ and $t = 3T/2$, when the pore width is
again larger than the static pore. Therefore, we observe a second peak in
the gain corresponding to $\tau_\mathrm{stat} \sim 3T/2$. This peak is
much smaller in magnitude because at higher frequencies a substantial amount of time
is spent by the polymer in a trapped state. As the frequency is increased further,
the gain becomes insensitive to the frequency and develops a plateau \cite{supp}.
Note that as the frequency tends to zero, the pore is essentially static and the polymer
translocation time reduces to the translocation time for the static pore and
the gain approaches unity. We have observed that the behavior of the system
is robust as the length of the polymer is changed. For example, Fig. \ref{f:sr}a shows
that for $N=128$ all the features are identical to the $N=32$ case, except that
two additional higher frequency peaks are observed.

We find that the gain increases as $\epsilon_\mathrm{pm}$ (that represents
the stickiness of the pore) is increased, although the general
features---the primary and secondary peaks at intermediate frequencies and
the saturation at high frequencies---are preserved (see Fig. \ref{f:sr}a).
While the low frequency limit of the gain is universal, we observe that
the saturation value of the gain at high frequencies depends on the external
force and the stickiness of the pore, as shown in Fig. \ref{f:sr}b.

\begin{figure}
\includegraphics[width=1.0\linewidth]{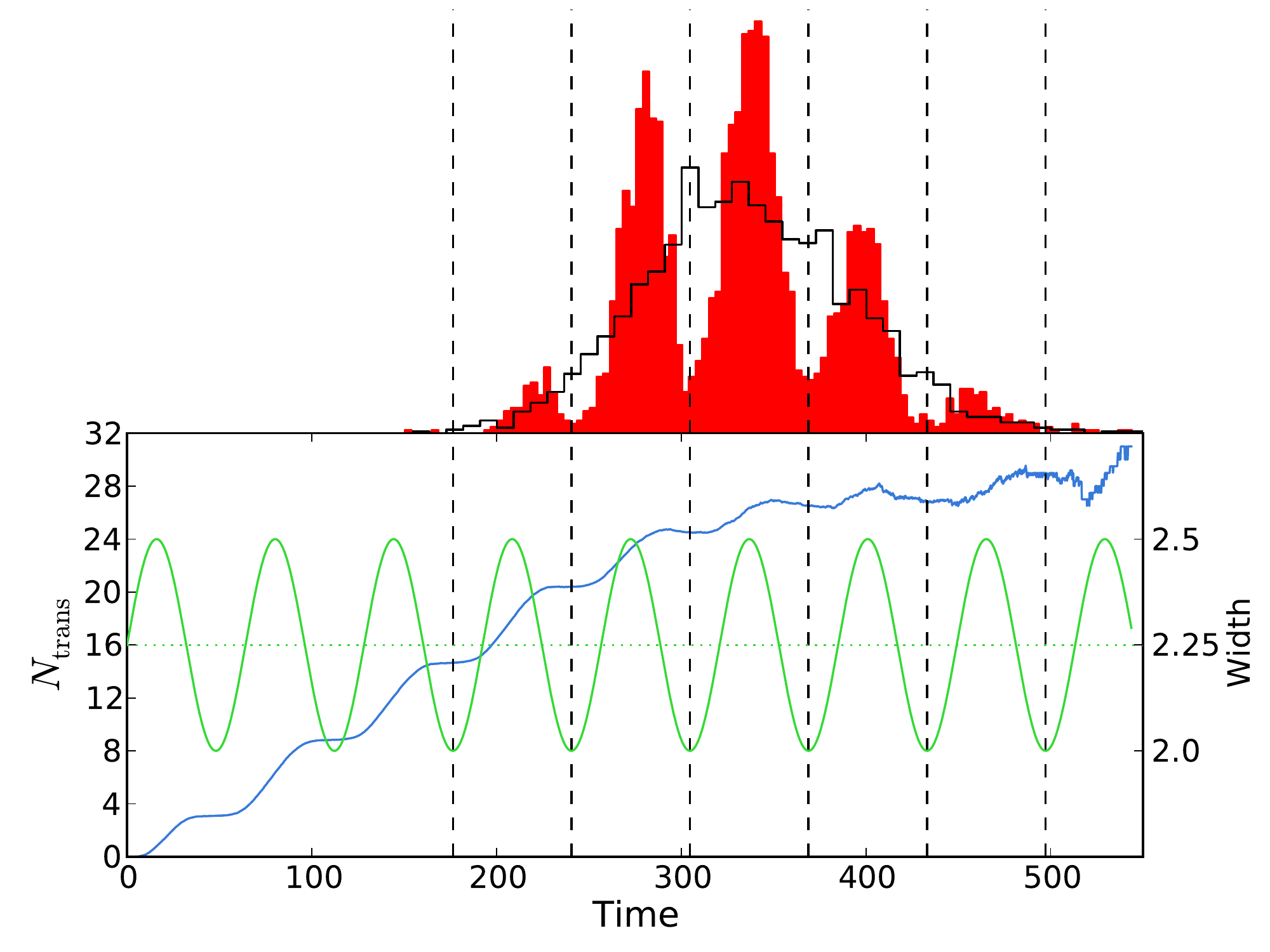}
\caption{\label{f:dist} Translocation time distribution (top)
as compared with the distribution for the static pore with average width (solid line), 
and average number of translocated monomers as a function of time (bottom)
at $\tilde{\omega} = 5.19$, for $F=1.0$ and $\epsilon_\mathrm{pm}=1.0$.
}
\end{figure}

Since the translocation is controlled by the constraint set by the time
varying width of the pore, studying how the translocation time for
a static pore depends on the width could help us understand the observed
behavior \cite{metzler2}. In Fig. \ref{f:sr}c, the average time for the translocation of
a polymer through a static pore is plotted as a function of the pore width.
One can identify two distinct behaviors, namely, a very sharp decrease
for $W_\mathrm{min}=2.0 < W < 2.25$, which crosses over to a regime with
relatively slower decay for $2.25 < W < L=5.0$. We find that the former
regime is controlled by the time the polymer takes to transfer through
the pore, whereas the latter is controlled by the time it takes to
escape the pore. The inset in Fig. \ref{f:sr}c shows the same behavior
for different values of the external force.

One can gain further insight by using a scaling argument for a confined
polymer within the blob picture \cite{sakaue2}. A confined polymer
that fills the entire pore of length $L$ breaks up into $L/W$ blobs of
uniform size $W$. However, during escape the polymer is split into
a fraction that fills a part of the pore of length $L-y$, leaving
the part of length $y$ empty, and a part that has gone outside.
The entropic penalty for the partial confinement of the polymer
inside the channel is $\sim k_{\rm B}T (L-y)/W$. The monomers
that are inside the channel will experience an external force $F$.
Therefore, displacing the polymer by a length $y$ inside the channel
will result in a gain of $-F(L^2-y^2+2 L y)/2W$ in mechanical energy.
The interaction between the monomers and the sticky walls of the pore
should also be taken into account. Since the LJ attraction is short ranged,
we can estimate this contribution by assigning an energy gain of
$-\epsilon_\mathrm{pm}$ to every monomer that is in the
vicinity of the channel walls. Counting these monomers, we find
this ``adsorption energy'' as
$-2\epsilon_\mathrm{pm}[(L-y)/W] W^{4/3}(1/W)$.
Therefore, the total free energy for escape (up to a constant) is given as
\begin{math}
{\cal F}(y)=\left[-\frac{c_1k_{\rm B}T}{W}
+ \frac{2c_2\epsilon_\mathrm{pm}}{W^{2/3}}
- \frac{F L}{W}\right]y + \frac{F y^2}{2W},
\end{math}
where $c_1$ and $c_2$ are constants of order unity.
We can now treat the escape problem as diffusion across a one-dimensional
effective potential barrier, and calculate the mean first passage time
\cite{lubensky}. The resulting escape time would lead to a similar trend as
in Fig. \ref{f:sr}c.

We can now calculate the translocation velocity, defined as
$V = N/\tau_{\mathrm{stat}}$, for a static pore as a function of the width,
and use the time dependence of the width 
to extract the translocation
velocity as a function of time over a full period. These are shown as insets in
Fig. \ref{f:sr}b. Integrating the velocity over a time range that is just long
enough to allow for the translocation of the entire polymer, we get the translocation
time for an oscillating pore within this approximate scheme. The solid lines
in Fig. \ref{f:sr}b show the result of this calculation for the high frequency saturation
value of the gain for different values of $F$ and $\epsilon_\mathrm{pm}$.
The instantaneous static pore picture thus seems to provide a reasonable account
of the phenomenon, although it systematically underestimates the gain due to the absence
of noise. The asymmetry observed in the two half cycles of the translocation
velocity is a direct consequence of choosing the average width $W_0$ to be
the crossover point between two different regimes in the translocation time versus
width plot in Fig. \ref{f:sr}c.

It is instructive to examine the distribution of the translocation times
in the high frequency regime. As shown in Fig. \ref{f:dist}, we find that
the distribution consists of a series of peaks separated by distinct minima.
These minima correspond to the points in the oscillation cycle when the
width of the pore is at its minimum, $W_\mathrm{min}$. We can further explore
the detailed dynamics of polymer translocation by looking at the number
of translocated monomers, $N_\mathrm{trans}$, i.e. the number of monomers
that have left the pore as a function of time. The plot shown in Fig. \ref{f:dist}
is calculated by averaging over all successful translocation events.
$N_\mathrm{trans}$ as a function of time alternates between flat and linearly
increasing regions. The flat regions correspond to the time periods
in the oscillation cycle when the pore width varies between $W_\mathrm{min}$
and $W_0$, when the polymer has very little space to manoeuvre itself
and is essentially trapped. Therefore, the monomers are not able to leave
the pore, and hence, $N_\mathrm{trans}$ does not change. During the periods
in the oscillation cycle when the width varies between $W_0$ and $W_\mathrm{max}$,
the polymer is no longer trapped and the monomers can escape
the pore resulting in an increase in $N_\mathrm{trans}$. Note that an increase
in the frequency of oscillation $\tilde{\omega}$ causes an
increase in the number of peaks in the translocation time distribution.

\begin{figure}
\includegraphics[width=1.0\linewidth]{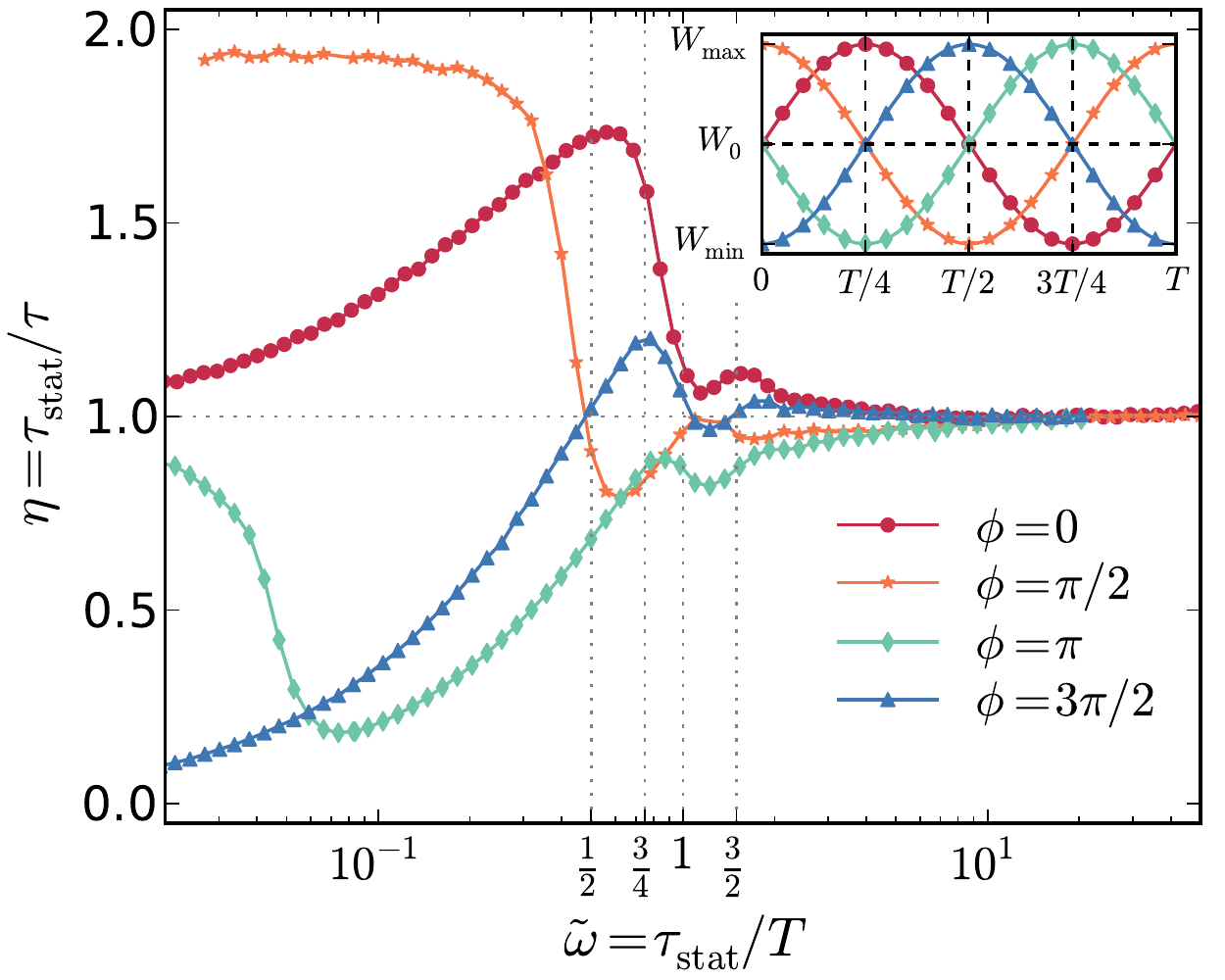}
\caption{\label{f:phase}
Gain as a function of the scaled frequency for different values of $\phi$.
The plots correspond to $F=1.0$ and $\epsilon_\mathrm{pm}=1.0$.
Inset shows the corresponding width oscillation cycles.}
\end{figure}

In our analysis so far we have assumed that the pore width at the beginning
of the translocation process is the mean width of the oscillation cycle.
Figure \ref{f:phase} shows the effect of the initial phase $\phi$ 
on the gain as a function of the oscillation frequency.
We observe that at low frequencies the initial phase strongly affects the translocation
gain by controlling the effective width during the translocation period, in agreement
with the picture described above. However, as the initial phase only affects the first
couple of cycles, we expect that at sufficiently high frequencies the translocation
process becomes insensitive to it. This is confirmed by the observation that the saturation
value of the gain at high frequencies is independent of the initial phase,
as Fig. \ref{f:phase} demonstrates. The robust high frequency behavior of the system
has remarkable implications. It suggest that even stochastic flickering of the pore
will result in a steady average translocation with a gain that can be tuned via $F$
and $\epsilon_\mathrm{pm}$, as long as $T \ll \tau_{\mathrm{stat}}$, where $T$ is
the characteristic period of the random opening and closing of the pore. Moreover,
we can infer from Fig. \ref{f:sr}b that the translocation rate for such random
flickering pores is extremely sensitive to the stickiness of the pore, which could
bring about the possibility of a high degree of robust selectivity in
the translocation process.

It will be interesting to probe whether such potential for selectivity is already
exploited in biological active pores. Other interesting factors could be
the flexibility of the polymer and the strength of the noise, which could potentially
lead to stochastic resonance of the polymer \cite{asfaw}. The design rules that can
be obtained from our study could also be used in fabricating highly selective
synthetic active pores, which might be able to perform tasks such as sequencing.

This work was supported by grant EP/G062137/1 from the EPSRC.


\begin{thebibliography}{99}

\bibitem{salman}
H. Salman et. al., Proc. Natl. Acad. Sci. USA {\bf 98}, 7247 (2001).

\bibitem{meller1}
A. Meller, J. Phys. Condens. Matter {\bf 15}, R581 (2003).

\bibitem{meller2}
A. Meller, L. Nivon, and D. Branton, Phys. Rev. Lett. {\bf 86}, 3435 (2001).

\bibitem{kasianowicz}
J. J. Kasianowicz, E. Brandin, D. Branton, and D.W. Deamer,
Proc. Natl. Acad. Sci. U.S.A. {\bf 93}, 13770 (1996).

\bibitem{akeson}
M. Akeson, D. Branton, J. J. Kasianowicz, E. Brandin, and
D.W. Deamer, Biophys. J. {\bf 77}, 3227 (1999).

\bibitem{sauer}
A. F. Sauer-Budge, J. A. Nyamwanda, D. K. Lubensky,
and D. Branton, Phys. Rev. Lett. {\bf 90}, 238101 (2003)

\bibitem{storm}
A. J. Storm, C. Storm, J. Chen, H. Zandbergen, J.-F.
Joanny, and C. Dekker, Nano Lett. {\bf 5}, 1193 (2005).

\bibitem{sung}
W. Sung and P. J. Park, Phys. Rev. Lett. {\bf 77}, 783 (1996).

\bibitem{muthukumar}
M. Muthukumar, J. Chem. Phys. {\bf 111}, 10371 (1999).

\bibitem{lubensky}
D. K. Lubensky and D. R. Nelson, Biophys. J. {\bf 77}, 1824 (1999).

\bibitem{metzler}
R. Metzler and J. Klafter, Biophys. J. {\bf 85}, 2776 (2003)

\bibitem{kantor}
J. Chuang, Y. Kantor, and M. Kardar, Phys. Rev. E {\bf 65}, 011802 (2001);
Y. Kantor and M. Kardar, Phys. Rev. E {\bf 69}, 021806 (2004).

\bibitem{milchev}
A. Milchev, K. Binder, and A. Bhattacharya, J. Chem. Phys. {\bf 121},
6042 (2004); A. Milchev, J. Phys. Condens. Matter {\bf 23}, 103101 (2011).

\bibitem{luo1}
K. Luo, T. Ala-Nissila, S. C. Ying, and A. Bhattacharya, Phys. Rev. Lett.
{\bf 99}, 148102 (2007); Phys. Rev. E {\bf 78}, 061918 (2008);
Phys. Rev. Lett. {\bf 100}, 058101 (2008); Phys. Rev. E {\bf 78},
061911 (2008).

\bibitem{luo2}
K. Luo, T. Ala-Nissila, S. C. Ying, and R. Metzler, Europhys. Lett.
{\bf 88}, 68006 (2009).

\bibitem{sakaue1}
T. Sakaue, Phys. Rev. E {\bf 76}, 021803 (2007); T. Saito and T. Sakaue
arXiv:1103.0620 (2011).

\bibitem{gerland}
U. Gerland, R. Bundschuh, and T. Hwa, Phys. Biol. {\bf 1}, 19 (2004).

\bibitem{slonkina}
E. Slonkina and A. B. Kolomeisky, J. Chem. Phys. {\bf 118}, 7112 (2003).

\bibitem{gopinathan}
A. Gopinathan and Y. W. Kim, Phys. Rev. Lett. {\bf 99}, 228106 (2007).

\bibitem{fazli}
N. Nikoofard and H. Fazli, Phys. Rev. E {\bf 83}, 050801 (R) (2011).

\bibitem{mitochondria}
P. Rehling, K. Brandner, and N. Pfanner, Nature Rev. Mol. Cell Biol. {\bf 5}, 519 (2004).

\bibitem{npc}
J. Yamada et al., Mol. Cell. Proteomics {\bf 9}, 2205 (2010).


\bibitem{supp}
See supplementary material at http://www-thphys.physics.ox.ac.uk/people/RaminGolestanian/
for animations showing the different translocation regimes at different frequencies.

\bibitem{metzler2}
K. Luo and R. Metzler, J. Chem. Phys. {\bf 134}, 135102 (2011).

\bibitem{sakaue2}
T. Sakaue and E. Raph\"{a}el, Macromol., {\bf 39} 2621 (2006).

\bibitem{asfaw}
M. Asfaw and W. Sung, Europhys. Lett. {\bf 90}, 30008 (2010).

\end{thebibliography}
\end{document}